\documentclass[12pt]{article}

\usepackage{graphicx}

\usepackage{scicite}

\usepackage{times}

\usepackage{hyperref}

\usepackage{amsmath,amssymb,latexsym}

\usepackage{color}

\def\msun{{\rm ~M}_{\odot}}
\def\rsun{{\rm ~R}_{\odot}}

\def\zsun{{\rm ~Z}_{\odot}}

\def\gpy{{\rm ~Gpc}^{-3} {\rm ~yr}^{-1}}

\newenvironment{sciabstract}{%
\begin{quote} \bf}
{\end{quote}}

\newcounter{lastnote}

\def\araa{{ARA\&A}}             
\def\apj{{Astrophys. J.}}                 
\def\apjl{{Astrophys. J. Lett.}}                
\def\apjs{{ApJS}}               
\def\apss{{Ap\&SS}}             
\def\aap{{Astron. Astrophys.}}                
\def\aapr{{A\&A~Rev.}}          

\def\mnras{{Mon. Not. R. Astron. Soc.}}             
\title{The first gravitational-wave source from the isolated evolution of
two $40$--$100\msun$ stars}

\author{
Krzysztof Belczynski,$^{1}$ Daniel E. Holz,$^{2}$ Tomasz Bulik,$^{1}$
Richard O'Shaughnessy,$^{3}$ 
\\
\normalsize{$^{1}$Astronomical Observatory, Warsaw University, Ujazdowskie 4, 00-478 Warsaw, Poland}\\
\normalsize{$^{2}$Enrico Fermi Institute, Department of Physics, Department of
  Astronomy \& Astrophysics, }\\
\normalsize{and Kavli Institute for Cosmological Physics, University of Chicago, Chicago, IL 60637, USA}\\
\normalsize{$^{3}$Center for Computational Relativity and Gravitation,}\\ 
\normalsize{Rochester Institute of Technology, Rochester, New York 14623, USA}
}

\date{}

\begin{document} 

\baselineskip24pt

\maketitle

\begin{sciabstract}
The merger of two massive $\sim 30 \msun$ black holes has been detected in
gravitational waves~\cite[GW150914]{DiscoveryPaper}. This discovery validates
recent predictions~\cite{Belczynski2010a,Dominik2015,Belczynski2015} that massive 
binary black holes would constitute the first detection. However, previous 
calculations have not sampled the relevant binary black hole progenitors---massive, 
low-metallicity binary stars---with sufficient accuracy and input physics to enable 
robust predictions to better than several orders of magnitude
~\cite{Tutukov1993,Lipunov1997,Nelemans2001,Voss2003,Belczynski2007,Mennekens2014}.
Here we report a suite of high-precision numerical simulations of binary
black hole formation via the evolution of isolated binary stars, providing
a framework to interpret GW150914 and predict the properties of subsequent
binary black hole gravitational-wave events.
Our models imply that these events form in an environment where the metallicity 
is less than 10 per cent of solar; have initial masses of $40$--$100\msun$; and 
interact through mass transfer and a common envelope phase. Their progenitors 
likely form either at $\sim2$ Gyr, or somewhat less likely, at $\sim11$ Gyr after 
the Big Bang. Most binary black holes form without supernova explosions, and 
their spins are nearly unchanged since birth, but do not have to be parallel.
The classical field formation of binary black holes proposed in this 
study, with low natal kicks and restricted common envelope evolution, produces 
$\sim 40$ times more binary black holes than dynamical formation channels 
involving globular clusters~\cite{Rodriguez2016} and is comparable to 
the rate from homogeneous evolution 
channels~\cite{Marchant2016,deMink2016,Eldridge2016,Woosley2016}.
Our calculations predict detections of $\sim 1,000$ black hole mergers per year 
with total mass of $20$--$80\msun$ once second generation ground-based 
gravitational wave observatories reach full sensitivity.
\end{sciabstract}

We study the formation of coalescing black hole binaries using the {\tt StarTrack} 
population synthesis code~\cite{Belczynski2002,Belczynski2008a}. This method has 
been updated to account for the formation of massive black hole systems in isolated 
stellar environments. The new key factors include observationally supported star 
formation rate, chemical enrichment across cosmic time and revised initial condition 
for evolution of binary stars.  
Hitherto, simulations have been unable to achieve the desired predictive power because
of the limitations on the input physics (e.g., limited metallicity range) and numerical 
accuracy. To ensure the dominant contribution from intrinsically rare low-metallicity
star-forming environments are adequately sampled, we employ a dense grid of
metallicities ($32$ metallicities) with high precision ($20$ million binaries each).

Although binary population synthesis is dependent on a number of uncertain physical 
factors, there has been recent progress in reducing this uncertainty and understanding 
how it affects predictions. In light of this we consider the following three models to 
encompass major sources of uncertainty~(Methods):
M1 represents our ``standard'' classical formation model for double compact objects 
composed of two black holes (BH-BH), two neutron stars (NS-NS), or one of each (BH-NS). 
M2 is our ``optimistic'' model in which Hertzsprung gap (HG) stars may initiate and 
survive common envelope (CE) evolution, leading to significantly more binaries being 
formed. M3 is our ``pessimistic'' model, where black holes receive high natal kicks, 
which disrupts and thereby reduces the number of BH-BH progenitor binaries.

For each generated double compact object merger, with its intrinsic component masses
and the redshift of the merger, we estimate the probability that such a merger would
have been detectable in the first observing run (O1) of Laser Interferometer 
Gravitational-Wave Observatory (LIGO) advanced detectors. We adopt a self-consistent 
model of evolution of stellar populations in Universe~\cite{Dominik2015,Belczynski2015}, 
and we take the representative noise curve for the O1 run~\cite{fn1}, and assume $16$ 
days of coincident science-quality observational time~\cite{DiscoveryPaper}.

\begin{figure}
\vspace*{-3.0cm}
\includegraphics[width=1.0\columnwidth]{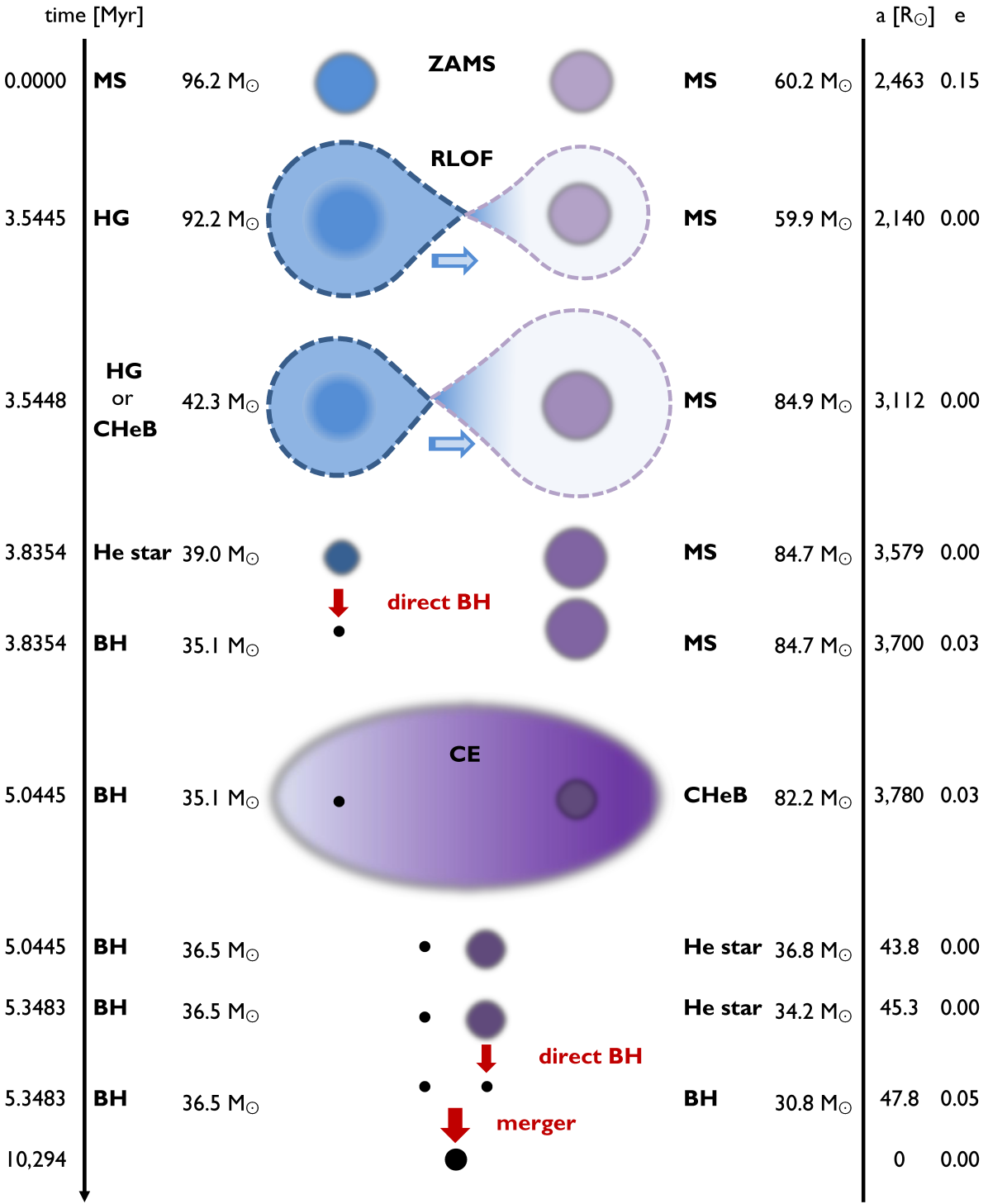}\vspace*{-1.0cm}\\
\caption{ 
{\bf Example binary evolution leading to a BH-BH merger similar to GW150914.}
A massive binary star ($96 + 60\msun$) is formed in the distant past ($2$ billion 
years after Big Bang; $z\sim3.2$) and after five million years of evolution forms 
a BH-BH system ($37 + 31\msun$). For the ensuing $10.3$ billion years this BH-BH 
system is subject to angular momentum loss, with the orbital separation steadily 
decreasing, until the black holes coalesce at redshift $z=0.09$. This example 
binary formed in a low metallicity environment ($Z=3\%\zsun$).}
\label{fig:ExamplEvol}
\end{figure}

In Figure~\ref{fig:ExamplEvol} we show the formation and evolution of a typical binary 
system which results in a merger with similar masses and at a similar time to GW150914. 
Stars that form such mergers are very massive ($40$--$100\msun$), and at the end of their 
lives they collapse directly to form BHs~\cite{Fryer2012}. Since there is no associated 
supernova explosion there is also no mass ejection. We allow $10\%$ of the collapsing 
stellar mass to be emitted in neutrinos. If natal kicks are associated with asymmetric 
mass ejection (as in our standard model), our prediction is that these massive BHs do not 
receive natal kicks and their spin directions are the same as that of their progenitor 
collapsing stars. The binary evolution removes the hydrogen-rich envelope from both 
binary components, making both stars compact and luminous Wolf-Rayet stars before they 
collapse to black holes. The first binary interaction is a dynamically stable Roche lobe 
overflow phase, while the second interaction consists of a common envelope phase that 
produces a compact binary. After the common envelope phase, the progenitor binary 
resembles two known high-mass X-ray binaries hosting massive black holes: IC10 X-1 and 
NGC300 X-1~\cite{Bulik2011}. A massive BH-BH binary (two $\sim 30\msun$ BHs) is formed 
in $\sim 5$ Myr of evolution, with a relatively wide orbit ($a\sim50\rsun$), leading to 
a long time to coalescence of $t_{\rm merger}\sim10$ Gyr. The accretion onto the first 
BH in the common envelope phase is only modest ($\Delta m \sim 1.5\msun$) while 
accretion from stellar wind of its companion is rather small ($\Delta m < 0.1\msun$).

To investigate general aspects of the formation history of GW150914, we select a 
population of ``GW150914-like'' BH-BH mergers with a total redshifted mass of 
$M_{\rm tot,z}=54$--$73\msun$, and then further restrict our sample to binaries that 
would be detectable in the first observing run (O1) of Laser Interferometer 
Gravitational-Wave Observatory (LIGO) advanced detectors. The formation channels typical 
for these massive BH-BH mergers are summarized in Extended Data Table~\ref{tab:EvolChan}. 

We find that the most likely progenitor of GW150914 consists of a primary star in the 
mass range $40$--$100\msun$ and a secondary in the mass range $40$--$80\msun$. In our 
standard model the binary formed in a low metallicity environment ($Z<10\%\zsun$; see 
Extended Data Fig.~\ref{fig:BHBHMass}) and in either the early Universe ($2$ Gyr after 
the Big Bang) or very recently ($11$ Gyr after the Big Bang).

\begin{figure}
\vspace*{-1.5cm}
\includegraphics[width=1.0\columnwidth]{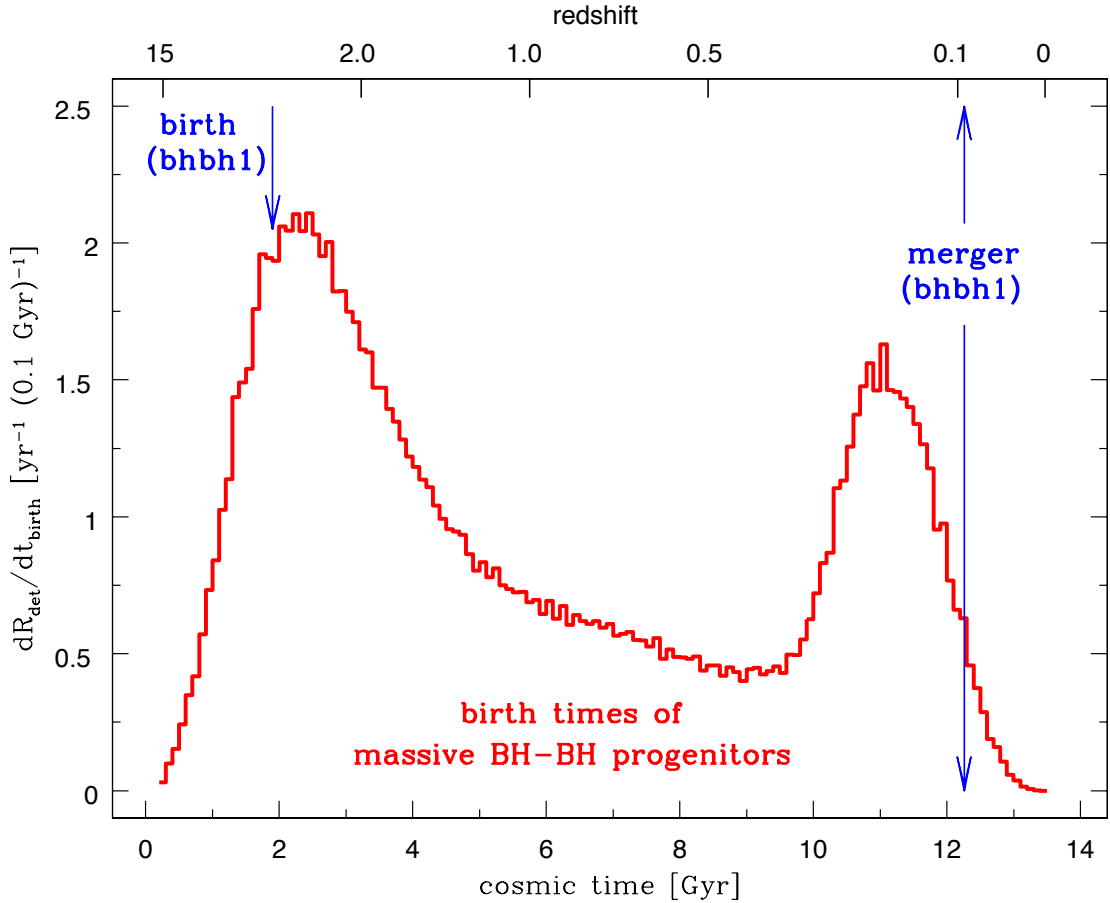}\vspace*{-4.6cm}\\
\caption{{\bf Birth times of GW150914-like progenitors across cosmic time.} 
Half of the binaries that form BH-BH mergers detectable in O1 with total redshifted 
mass in the range $M_{\rm tot,z}=54$--$73\msun$  were born within $4.7$~Gyr of the 
Big Bang (corresponding to $z>1.2$). The birth and merger times of binary from 
Figure~\ref{fig:ExamplEvol} is marked; it follows the most typical evolutionary 
channel for massive BH-BH mergers (BHBH1 in Extended Data Tab.~\ref{tab:EvolChan}). 
Note that the merger redshift of GW150914 is $z=0.088$. The bimodal shape of the 
distribution originates from a combination of the BH-BH delay time distribution 
with the low-metallicity star formation history (Extended Data Fig.~\ref{fig:Emergence} 
for details).}
\label{fig:BirthTimes}
\end{figure}

The distribution of birth times of these massive BH-BH mergers is bimodal 
(Fig.~\ref{fig:BirthTimes} and Extended Data Fig.~\ref{fig:Emergence}), with a majority 
of systems originating from the distant past ($55\%$ of binaries; $\sim 2$~Gyr after the 
Big Bang corresponding to $z\sim3$), and a smaller contribution from relatively young 
binaries ($25\%$; formed $\sim 11$~Gyr after the Big Bang corresponding to $z\sim0.2$). 
This bimodality arises from two naturally competing effects: On the one hand, most 
low-metallicity star formation occurs in the early Universe. On the other hand, in 
contrast to our prior work~\cite{Dominik2015,Belczynski2015}), significantly more 
low-metallicity star formation is currently expected to occur in the low-redshift 
Universe~\cite{Hirschauer2016}. 
Therefore, as is the case with binary neutron stars, we anticipate a 
significant contribution to the present-day binary black hole merger rate from 
binary black holes formed in low-redshift low-metallicity star forming regions. 
The delay time distribution of BH-BH binaries in our simulations follows a
$1/t$ distribution. The birth times therefore naturally pile up at low redshifts
($z \sim 0.1$--$0.3$)  and this gives rise to a low-$z$ peak (Extended Data
Fig.~\ref{fig:Emergence}; a). 
However, the low-metallicity star formation ($Z<10\%\zsun$) responsible for the 
production of massive BH-BH mergers peaks at redshift $\sim 3$ (Extended Data
Fig.~\ref{fig:Emergence}; b). The convolution of these two effects 
produces the bimodal birth time distribution (Extended Data Fig.~\ref{fig:Emergence};
c). 

These massive ``GW150914-like'' mergers consist of comparable mass black holes. The 
vast majority ($99.8\%$) of mergers are found with mass ratios in the range 
$q=0.7$--$1.0$ (Extended Data Fig.~\ref{fig:MassRatio}), with the mass ratio of 
GW150914 ($q=0.82_{-0.21}^{+0.16}$) falling near the center of the expected  region. 
The formation of low mass ratio objects is suppressed because low mass ratio progenitors 
tend to merge during the first mass transfer event when the more massive component 
overfills its Roche lobe~\cite{Bulik2004}. However, with decreasing total merger mass, 
the mass ratio extends to lower values. In particular, for the lower mass bin of 
$M_{\rm tot,z}=25$--$37\msun$, mass ratios as low as $q=0.3$ are also found. 

\begin{figure}
\vspace*{-7.0cm}
\hspace*{-0.8cm}
\includegraphics[width=1.1\columnwidth]{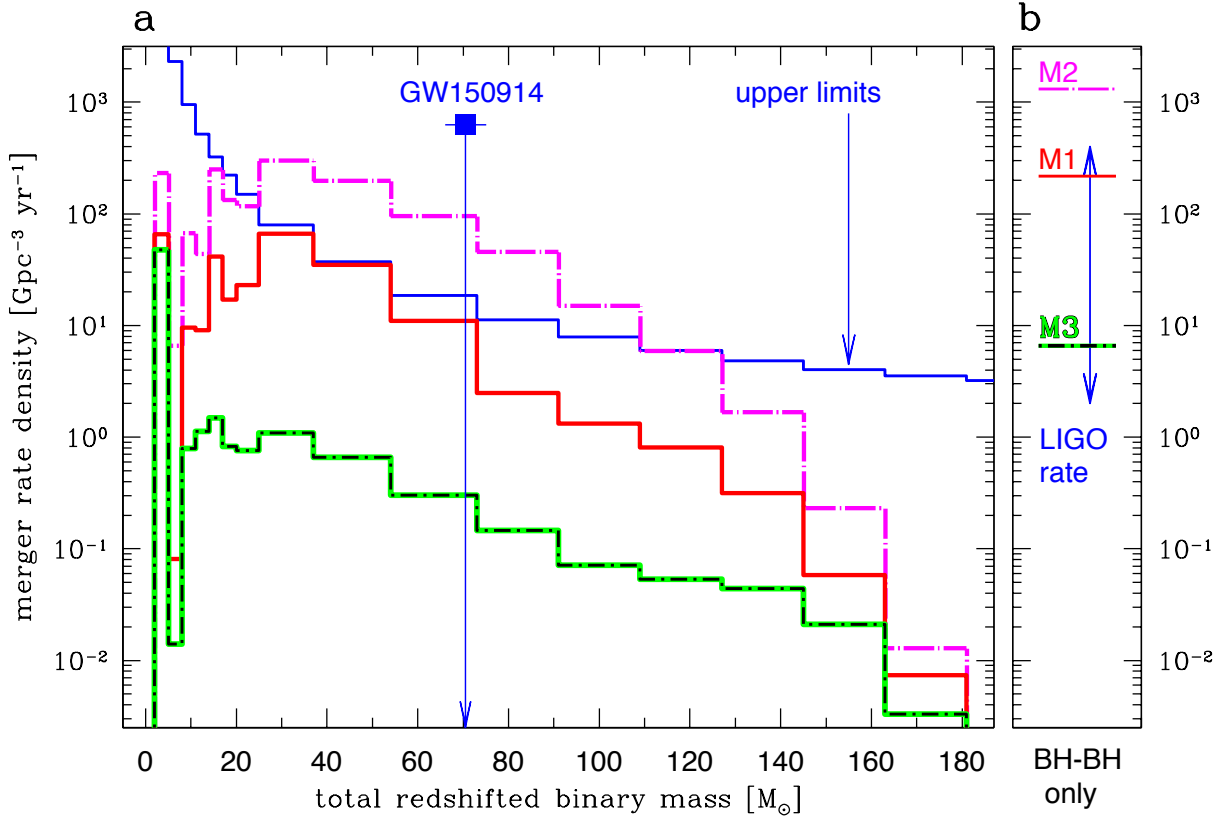}\vspace*{-5.0cm}\\
\caption{{\bf Comparison of merger rates and masses with LIGO O1 results:} 
for standard (M1), optimistic CE (M2), and pessimistic high BH kicks (M3) models.    
{\bf a,} Total redshifted binary merger mass distribution. GW150914 
($70.5\msun$: blue square with $90\%$ confidence interval in mass). The blue line 
shows the fiducial estimate of the sensitivity of the 16 day O1 run.
A comparison of the shapes of the blue and red curves suggests that the most 
likely detections for M1 are BH-BH mergers with mass in the range $25$--$73\msun$. 
NS-NS mergers (first bin) and BH-NS mergers (next five bins) are well below the    
estimated sensitivity, and thus detections in O1 are not expected. The rate 
densities  are in the detector rest frame. 
{\bf b,} Comparison of the LIGO BH-BH rate estimate with our models.
The LIGO value of $2$--$400\gpy$ ($90\%$ credible range) compares well with our
standard and high BH natal kick models. The rate densities  are in the source 
rest frame.
{\bf Updated version of this figure with the most recent LIGO observations
may be found at:} \url{www.syntheticuniverse.org/stvsgwo.html}
}
\label{fig:RateComp}
\end{figure}

We now use our full sample of double compact object mergers to make predictions for 
the merger rate density, detection rates, and merger mass distribution. The results 
are shown in Figure~\ref{fig:RateComp} and Extended Data Table~\ref{tab:EventRate}, 
where we compare them to the measured values inferred from LIGO O1 observations. We 
find an overall detection rate consistent with the detection of one significant 
candidate (GW150914) during the principal $16$ day double-coincident period for our 
``standard'' model (M1), while it is inconsistent for our other two models 
(``optimistic'' M2 and ``pessimistic'' M3; more detail below). 

The BH-BH rates inferred from the 16 days of LIGO O1 observations are in the range 
$2$--$400$ Gpc$^{-3}$ yr$^{-1}$~\cite{RatesPaper}. 
For comparison, we estimate the rate density of binary black holes from our 
population synthesis data set. We consider the full population of binary black 
holes within a redshift of $z=0.1$ (i.e., not weighted by their detection 
probability), and calculate their average source-frame merger rate density.
We find a value of $218$ Gpc$^{-3}$ yr$^{-1}$ for our standard model (M1), which 
is in good  agreement with the inferred LIGO rate~\cite{RatesPaper}. In contrast,
our optimistic model (M2) predicts too many mergers, with a rate density
of $1,303$ Gpc$^{-3}$ yr$^{-1}$, while our pessimistic model (M3) is at the very
bottom end of the allowable range with a predicted rate of $6.6$ Gpc$^{-3}$
yr$^{-1}$. In our models, the BH-BH merger rate density increases with redshift 
(Extended Data Fig.~\ref{fig:IntrRate}). This increase is modest; our predicted 
source-frame BH-BH merger rate density would double if the cutoff redshift was 
increased from $z=0.1$ to $z=0.6$. 

The merger rate density for model with optimistic common envelope (M2) is an 
order of magnitude larger than the rate estimate from LIGO. This implies that 
unevolved massive stars (during main sequence and Hertzsprung gap) do not
initiate/survive CE~\cite{Pavlovskii2015,Belczynski2007}. In our classical 
BH-BH formation scheme only evolved stars (during core helium burning) with 
well-developed convective envelopes are allowed to initiate and survive CE.

Our predictions for the pessimistic model (M3) imply that large natal kicks (with 
average magnitude $\gtrsim 400$ km s$^{-1}$) are unlikely for massive black 
holes. This model predicts that an event like GW150914 would happen only $1\%$ 
of the time, with the detection of any BH-BH system happening less that $10\%$ 
of the time (Tab.~\ref{tab:EventRate}). 
In principle this conclusion applies only to the formation of the first BH in the 
binary, since high natal kicks lead to disruption of BH-BH progenitors while the 
binaries are wide. During the formation of the second BH the progenitor binaries 
are on very close orbits (Fig.~\ref{fig:ExamplEvol}) and are not disrupted by natal 
kicks. In Extended Data Figure~\ref{fig:IntrRate} we show a sequence of models 
with intermediate BH natal kicks; future observations may allow us to discriminate 
between these models and constrain the natal kick distribution. If future 
observations converge on M1 it will indicate no natal kicks nor supernova 
explosions in massive BH formation~\cite{Fryer2012}. A striking ramification of 
this is the prediction that hot and luminous Wolf-Rayet progenitors of massive 
BHs~\cite{Eldridge2013} should disappear from the sky as a result of direct 
collapse to a BH. Targeted observational campaigns to search for such phenomena 
are already underway~\cite{Gerke2015}.

Figure~\ref{fig:RateComp} shows the relative contribution to the overall merger rate 
density associated with each bin of total redshifted merger mass $M_{\rm tot,z}$. 
For comparison, this Figure also shows the fiducial sensitivity ($0.7/VT$) as a 
function of mass, assuming equal-mass zero-spin binary black holes. This Figure 
demonstrates that the intersection of the strongly mass-dependent sensitivity and 
the intrinsic detectable mass distribution strongly favors sources with total 
redshifted masses between $25$--$73\msun$, consistent with our recent 
work~\cite{Belczynski2015}, and matching the total redshifted mass of GW150914 
($M_{\rm tot,z}=70.5\msun$). In our simulations the maximum intrinsic mass of a 
merging BH-BH binary is $M_{\rm tot}=140\msun$. When accounting for cosmological 
redshift ($M_{\rm tot,z} =(1+z)M_{\rm tot}$), and taking into account the advanced 
LIGO O1 horizon redshift for this most massive binary ($z=0.7$), the highest possible 
observed mass within the O1 run would be $\approx 240$M$_\odot$. 

Spin magnitudes and directions of merging black holes are potentially measurable by 
LIGO~\cite{DiscoveryPaper}.  
The second-born BH in a BH-BH binary does not accrete mass, and its spin at merger is 
unchanged from its spin at birth. The first-born BH, on the other hand, has a chance to 
accrete material from the unevolved companion's stellar wind or during CE evolution. 
However, since this is limited either by the very low efficiency of accretion from stellar 
winds or by inefficient accretion during CE~\cite{Ricker2008,Macleod2015}, the total 
accreted mass onto the first-born BH is expected to be rather small ($\sim 1$--$2\msun$). 
This is insufficient to significantly increase the spin, and thus the first-born BH spin 
magnitude at merger is within $\sim 10\%$ of its birth spin. 

In our modeling we {\em assume}\/ that stars that are born in a binary have their
spins aligned with the binary angular momentum vector. If massive black holes do not 
receive natal kicks (e.g., our standard model M1), then our prediction is that BH spins 
are aligned during the final massive BH-BH merger. We note that our standard model 
includes natal kicks and mass loss for low-mass BHs ($\lesssim 10\msun$), and therefore 
BH-BH binaries with one or two low-mass BHs may show misalignment.
Alternatively, binaries could be born with misalignment and retain it, or misalignment 
could be caused by the third body, or by interaction of radiative envelope with convective 
core~\cite{Rogers2013}, or misalignment could result from high natal kick on the 
second-born BH. Several binaries are reported with misaligned spins~\cite{Albrecht2014}.
Therefore, spin alignment of massive merging black holes suggests isolated field evolution, 
while misaligned spins do not elucidate formation processes. 

As shown in Figure~\ref{fig:ExamplEvol}, we find that the formation of massive BH-BH 
mergers is a natural consequence of isolated binary evolution. Our standard model (M1) 
of BH-BH mergers fully accounts for the observed merger rate density and merger mass 
(Fig.~\ref{fig:RateComp}) and mass ratio of two merging BHs (Extended Data 
Fig.~\ref{fig:MassRatio}) inferred from GW150914. 

Our standard formation mechanism (M1) produces significantly more binary black 
holes than alternative, dynamical channels associated with  globular clusters.   
A recent study~\cite{Rodriguez2016} suggested globular clusters could produce a 
typical merger rate of $5$ Gpc$^{-3}$ yr$^{-1}$; our standard (M1) model BH-BH 
merger rate density is $\sim 40$ times higher: $218$ Gpc$^{-3}$ yr$^{-1}$.  

However, one non-classical isolated binary evolution channel involving rapidly 
rotating stars (homogeneous evolution) in very close binaries may also fully account 
for the formation of GW150914~\cite{Marchant2016,deMink2016,Eldridge2016,Woosley2016}. 
In particular, typical rates of 1.8 detections in 16 days of O1 observations are 
found~\cite{deMink2016}, which is comparable to our prediction of 2.8. 
(Tab.~\ref{tab:EventRate}). 
Only very massive BH-BH mergers with total intrinsic mass $\gtrsim 50\msun$ are 
formed in this model\cite{Marchant2016,deMink2016}, while we note that our model 
predicts mergers with mass in a broader range down to $\gtrsim 10\msun$. Future LIGO 
observations of BH-BH mergers may allow us to discriminate between these two very 
different mass distributions/models.

\bibstyle{Science}
\bibstyle{naturemag}

\newcounter{firstbib}
\bibliographystyle{naturemag}

\noindent
{\bf Acknowledgments.} 
We are indebted to Grzegorz Wiktorowicz, Wojciech Gladysz and Krzysztof Piszczek for 
their help with population synthesis calculations, and to Hsin-Yu Chen and Zoheyr 
Doctor for their help with our LIGO/Virgo rate calculations.  
We would like to thank thousands of {\tt Universe@home} users that have provided 
their personal computers for our simulations. We also want to thank the Hannover GW 
group for letting us use their {\tt ATLAS} supercomputer. 
KB acknowledges support from the NCN grant Sonata Bis 2 (DEC-2012/07/E/ST9/01360).
DEH was supported by \textsc{nsf career} grant PHY-1151836. He also acknowledges support
from the Kavli Institute for Cosmological Physics at the University of Chicago through 
\textsc{nsf} grant PHY-1125897 as well as an endowment from the Kavli Foundation.
TB acknowledges support from the NCN grant Harmonia 6 (UMO-2014/14/M/ST9/00707).
ROS was supported by \textsc{nsf} grant PHY-1505629.

\noindent
{\bf Author Contributions.} 
All authors contributed to the analysis and writing of the paper.

\noindent
{\bf Author Information.} 
Corresponding author email address: chrisbelczynski@gmail.com

\newpage


\begin{table}
\begin{center}
\begin{tabular}{cl|ccc}
Model& Type & O1 rate [yr$^{-1}$] & O1: 16 days\\ 
\hline \hline
    & All      & 63.18 & 2.770 \\
    & NS-NS    & 0.052 & 0.002 \\
M1  & BH-NS    & 0.231 & 0.010 \\
    & BH-BH    & 62.90 & 2.758 \\
    & GW150914 & 11.95 & 0.524 \\
\hline
    & All      & 476.1 & 20.87 \\
    & NS-NS    & 0.191 & 0.008 \\
M2  & BH-NS    & 0.796 & 0.035 \\
    & BH-BH    & 475.1 & 20.83 \\
    & GW150914 & 110.0 & 4.823 \\
\hline
    & All      & 1.985 & 0.087 \\
    & NS-NS    & 0.039 & 0.002 \\
M3  & BH-NS    & 0.014 & 0.001 \\
    & BH-BH    & 1.932 & 0.085 \\
    & GW150914 & 0.270 & 0.012 \\
\hline
\end{tabular}
\caption{\label{tab:EventRate}{\bf Expected detection rate and number of 
detections}: 
The first column marks: standard (M1), optimistic common envelope (M2), and 
high BH kicks (M3) models. 
The third column lists the expected detection rate $R_{\rm d}$ per unit 
double-coincident time (both LIGO detectors operating at appropriate 
sensitivity), for a network comparable to O1, shown for different classes of 
mergers. The fourth column shows $R_{\rm d} T$, where $T=16$ days is the 
analysis time relevant for the rate estimate for GW150914~\cite{RatesPaper}.
Entries marked with ``GW150914'' are for the subpopulation of BH-BH
mergers with total redshifted mass in the range $M_{\rm tot,z}=54$--$73\msun$. 
}
\end{center}
\end{table}
\clearpage

\clearpage

\vspace*{2cm}
\section*{The Methods}

Our Monte Carlo evolutionary modeling is performed with the {\tt StarTrack} binary 
population synthesis code~\cite{Belczynski2002}. In particular, we incorporate a 
calibrated treatment of tidal interactions in close binaries~\cite{Belczynski2008a}, 
a physical measure of the common envelope (CE) binding energy~\cite{Dominik2012,Xu2010}, 
and a rapid explosion supernova model that reproduces the observed mass gap between 
neutron stars and black holes~\cite{Fryer2012,Belczynski2012}. Our updated mass spectrum 
of black holes shows a strong dependence on the metallicity of the progenitor stars 
(Extended Data Fig.~\ref{fig:BHMassRel}). In galaxies with metallicities similar to the 
Milky Way ($Z=\zsun=0.02$) black holes formed out of single massive stars (initial mass 
$M_{\rm ZAMS}=150\msun$) reach a maximum mass of $M_{\rm BH}=15\msun$, while for very 
low metallicity ($Z=0.0001=0.5\%\zsun$) the maximum mass becomes $M_{\rm BH}=94\msun$. 
The above input physics represents our standard model (M1), which is representative of 
our classical formation scheme for double compact objects (BH-BH, BH-NS, and NS-NS). 

We have adopted specific values for a number of evolutionary 
parameters. Single stars are evolved with calibrated formulae based on detailed 
evolutionary calculations~\cite{Hurley2000}. Massive star winds are adopted from 
detailed studies of radiation driven mass loss~\cite{Vink2011}. For the Luminous Blue 
Variable phase the high mass loss rate is adopted ($1.5 \times 10^{-4} \msun$ yr$^{-1}$).
Binary interactions, and in particular the stability of RLOF, is judged based on
binary parameters: mass ratio, evolutionary stage of donor, response to mass loss, and 
behavior of the orbital separation in response to mass transfer. The orbital separation  
is additionally affected by gravitational radiation, magnetic braking, and angular 
momentum loss associated with systemic mass loss. During stable RLOF we assume that 
half of the mass is accreted onto the companion, while the other half ($1-f_{\rm a}=0.5$) 
is lost with specific angular momentum ($dJ/dt= j_{\rm loss} 
[J_{\rm orb}/(M_{\rm don}+M_{\rm acc})] (1-f_{\rm a}) dM_{\rm RLOF}/dt$ with 
$j_{\rm loss}=1.0$~\cite{Podsiadlowski1992}). 
CE is treated with energy balance with fully effective conversion of orbital energy into 
envelope ejection ($\alpha=1.0$), while the envelope binding energy for massive stars 
is calibrated by a parameter $\lambda$ that depends on star radius, mass, and metallicity. 
For massive stars $\lambda \approx 0.1$ is adopted~\cite{Xu2010}. 
During CE compact objects accrete at $10\%$ Bondi-Hoyle rate as estimated by recent 
hydrodynamical simulations~\cite{Ricker2008,Macleod2015}. 
Our CE is done ``instantaneously'', so the time at the beginning
and end of CE is exactly the same (see Fig.~\ref{fig:ExamplEvol}); the
time duration of CE has no impact on our results.

We consider two extra variations of the binary evolution input physics. In one model 
(M2) we test highly uncertain CE physics~\cite{Ivanova2013} and we allow for 
Hertzsprung gap (HG) stars to initiate and survive CE evolution. This is an 
optimistic assumption, since these stars may not allow for CE 
evolution~\cite{Pavlovskii2015}, nor survive as a binary if CE does 
happen~\cite{Belczynski2007}.
For comparison, in our standard model we allow only evolved stars with a deep
convective envelope (core Helium burning stars) to survive CE.  

In the opposite extreme, we employ a model (M3) where black holes receive high 
natal kicks. In particular, each BH gets a natal kick with its 
components drawn from a $1$-D Maxwellian distribution with $\sigma=265$ km
s$^{-1}$, independent of BH mass. 
Such high natal kicks are measured for Galactic pulsars~\cite{Hobbs2005}. 
This is a pessimistic assumption, as high natal kicks tend to disrupt BH-BH
progenitor binaries. This assumption is not yet excluded based on
electromagnetic observations~\cite{Belczynski2015}.  
In contrast, in our standard model BH natal kicks decrease with BH mass. In 
particular, for massive BHs that form through direct collapse of an entire 
star to a BH with no supernova explosion
($M_{\rm BH} \gtrsim 10\msun$ for solar metallicity; 
$M_{\rm BH} \gtrsim 15\msun$ for $Z=10\%\zsun$; and $M_{\rm BH} \gtrsim 15$--$30\msun$ 
for $Z=1\%\zsun$) we assume no natal kicks~\cite{Fryer2012}.
We have also calculated a series of models with intermediate BH kicks (see Extended Data 
Fig.~\ref{fig:IntrRate}):
\mbox{$\sigma=200$ km\ s}$^{-1}$ (model M4), 
$\sigma=130$ km\ s$^{-1}$ (model M5), 
$\sigma=70$ km\ s$^{-1}$ (model M6).

For each evolutionary model we compute $2 \times 10^7$ massive binaries for each
point on a grid of $32$ sub-models covering a wide range of metallicities: 
$Z=0.0001$, $0.0002$, $0.0003$, $0.0004$, $0.0005$, $0.0006$, $0.0007$, $0.0008$, 
$0.0009$, $0.001$, $0.0015$, $0.002$, $0.0025$, $0.003$, $0.0035$, $0.004$, $0.0045$, 
$0.005$, $0.0055$, $0.006$, $0.0065$, $0.007$, $0.0075$, $0.008$, $0.0085$, $0.009$, 
$0.0095$, $0.01$, $0.015$, $0.02$, $0.025$, $0.03$.  
We assume that stellar evolution at even lower metallicities proceeds in the same way 
as the evolution at $Z=0.5\%\zsun$. However, we note that stars with very low metal 
content (e.g., Population III) may evolve differently than metal-rich 
stars~\cite{Szecsi2015}.

Each sub-model is computed with initial distributions of orbital periods
($\propto (\log P)^{-0.5}$), eccentricities ($\propto e^{-0.42}$), and mass ratios 
($\propto q^{0}$) appropriate for massive stars~\cite{Sana2012}.
We adopt an initial mass function that is close to flat for low mass stars
($\propto M^{-1.3}$ for $0.08 \leq M<0.5\msun$ and $\propto M^{-2.2}$ for $0.5 
\leq M<1.0\msun$) and top heavy for massive stars ($\propto M^{-2.3}$ for $1.0 \leq 
M \leq 150\msun$), as guided by recent observations~\cite{Bastian2010}. The adopted 
IMF generates higher BH-BH merger rate densities as compared with the steeper IMF 
($\propto M^{-2.7}$ for $1.0 \leq M \leq 150\msun$) adopted in our earlier 
studies~\cite{Dominik2013,Belczynski2015} as there are more BH-BH merger 
progenitors in our simulations~\cite{deMink2015}.

A moderate binary fraction ($f_{\rm bi}=0.5$) is adopted for stars with mass 
$M_{\rm ZAMS}<10\msun$, while we assume that all more massive stars are formed in 
binaries ($f_{\rm bi}=1.0$) as indicated by recent empirical 
estimates~\cite{Sana2012,Duchene2013}).

We adopt an extinction corrected cosmic star formation rate based on numerous 
multi-wavelength observations~\cite{Madau2014}:
\begin{equation}
\mbox{SFR}(z)=0.015 {(1+z)^{2.7} \over 1+[(1+z)/2.9]^{5.6}} \,\msun \, {\rm Mpc}^{-3} 
\, {\rm yr}^{-1}.
\label{sfr}
\end{equation} 
This SFR declines rapidly at high redshifts ($z>2$). This may be contrasted with some 
SFR models that we have used in the past~\cite{Strolger2004} which generated a greater 
number of stars at high redshifts. This revision will thus reduce the BH-BH merger 
rate densities at {\em all} redshifts. Even though the formation of BH-BH binaries 
takes a very short time ($\sim 5$ Myr), the time to coalescence of two black holes 
may take a very long time (Fig.~\ref{fig:ExamplEvol} and Extended Data 
Fig.~\ref{fig:Emergence}).

In our new treatment of chemical enrichment of the Universe we follow the mean 
metallicity increase with cosmic time (since Big Bang till present). The mean 
metallicity as a function of redshift is given by
\begin{equation}
\log(Z_{\rm mean}(z))=0.5+\log \left(  {y \, (1-R) \over \rho_{\rm b}} 
\int_z^{20} {97.8\times10^{10} \, sfr(z') \over H_0 \, E(z') \, (1+z')} dz' \right)
\label{Zmean}
\end{equation}
with a return fraction $R=0.27$ (mass fraction of each generation of stars that 
is put back into the interstellar medium), a net metal yield $y=0.019$ (mass of new 
metal created and ejected into the interstellar medium by each generation of stars 
per unit mass locked in stars), a baryon density $\rho_{\rm b}=2.77 \times 10^{11} 
\,\Omega_{\rm b}\,h_0^2\,\msun\,{\rm Mpc}^{-3}$ with $\Omega_{\rm b}=0.045$ 
and $h_0=0.7$, a star formation rate given by eq.~\ref{sfr}, and
$E(z)=\sqrt{\Omega_{\rm M}(1+z)^3+\Omega_{\rm k}(1+z)^2+\Omega_\Lambda)}$
with $\Omega_\Lambda=0.7$, $\Omega_{\rm M}=0.3$, $\Omega_{\rm k}=0$,
and $H_0=70.0\,{\rm km}\,{\rm s}^{-1}\,{\rm Mpc}^{-1}$. 
The shape of the mean metallicity dependence on redshift follows recent 
estimates~\cite{Madau2014}, although the level was increased by $0.5$ dex to better 
fit observational data~\cite{Vangioni2015}.
At each redshift we assume a log--normal distribution of metallicity around the 
mean, with $\sigma=0.5$ dex~\cite{Dvorkin2015}. 
Our prescription (Extended Data Fig.~\ref{fig:MetEvol}) produces more low-metallicity 
stars than previously~\cite{Dominik2013}. Since BH-BH formation is enhanced at
low-metallicity~\cite{Belczynski2010a}, our new approach increases the predicted rate 
densities of BH-BH mergers.

Here we discuss caveats of evolutionary calculations. 
First, we only consider isolated binary evolution, and thus our approach is
applicable to field stars in low density environments. It is possible that 
dynamical interactions enhance BH-BH merger formation in dense globular
clusters~\cite{Rodriguez2016}, offering a completely independent channel.

Second, our predictions are based on a ``classical'' theory of stellar and
binary evolution for the modeling of massive stars that we have compiled,
developed, and calibrated over the last $15$ years. We do not consider exotic
channels for the formation of BH-BH mergers, such as the one from rapidly 
rotating stars in contact binaries~\cite{Almeida2015}. 

Third, our modeling includes only three evolutionary models: a ``standard'' 
model consisting of our best estimates for reasonable parameters, as well as
``optimistic'' and ``pessimistic'' alternate models. The optimistic model
consists of only one change from the standard model: we allow all stars beyond
the main sequence to survive the common envelope phase. Alternatively, the
pessimistic model also consists of only one change: larger BH natal kicks. 
We have not investigated other possible deviations from the standard model (e.g., 
different assumptions of mass and angular momentum loss during stable mass 
transfer evolution) nor have we checked inter-parameter degeneracies (e.g., 
models with high BH kicks {\em and}\/ an optimistic common envelope phase). 
Albeit with low statistics and limited scope, precursor versions of these 
computationally demanding studies have already been performed~\cite{O'Shaughnessy2007}; 
these calculations indicate that our three models are likely to cover the range 
of interesting effects.

Fourth, our observations are severely statistically limited. We are attempting 
to draw inferences about our models based on a single detection (GW150914). 

In was argued~\cite{AstroPaper} that the formation of GW150914 in isolated 
binary evolution requires a metallicity lower than $50\%\zsun$. This was based on 
single stellar models~\cite{Spera2015}; stars in close binaries are subject to 
significant mass loss during RLOF/CE, and they form BHs with lower mass than single 
stars. Thus in binaries the metallicity threshold for massive BH formation is lower 
than in single stellar evolution. For example, formation of a {\em single}\/ $30\msun$ 
BH requires $Z<25\%\zsun$ (Extended Data Fig.~\ref{fig:BHMassRel}), while formation 
of two such BHs in a {\em binary} requires $Z<10\%\zsun$ (Extended Data 
Fig.~\ref{fig:BHBHMass}). The value of this threshold depends on assumptions 
for the model of stellar evolution, winds, and BH formation processes. The physical 
models we have adopted yield a threshold of $Z<10\%\zsun$, the same as the one 
obtained with MESA for homogeneous stellar evolution~\cite{Marchant2016}.
Our model was calibrated on known masses of BHs, and in particular we do not 
exceed $15\msun$ for $\zsun$ (the highest mass stellar BH known in our Galaxy). 
In contrast, single stellar models used to derive the high metallicity threshold 
produce $25\msun$ for $\zsun$~\cite{Spera2015}. The highest threshold obtained 
with binary evolution was reported at the level of $50\%\zsun$~\cite{Eldridge2016}. 
Such high value of metallicity threshold for the progenitor of GW150914 implies that 
stars at approximately solar metallicity ($Z=0.014$) produce BHs as massive as 
$40\msun$~\cite{Eldridge2016}. This is neither supported nor excluded by available 
electromagnetic BH mass measurements~\cite{fn2}. 

In the following we present calculation of the gravitational radiation signal.  
The output of {\tt StarTrack} is a binary merger at a given time. We then
calculate the gravitational waveform associated with this merger, and determine
whether this binary would have been observable by LIGO in the O1
configuration~\cite{Dominik2015,Belczynski2015}. 

We model the full inspiral-merger-ringdown waveform of the binaries using the 
IMRPhenomD gravitational waveform template family~\cite{Khan2015,Husa2015}. This
is a simple and fast waveform family which neglects the effects of spin (which
are not relevant for GW150914). We consider a detection to be given by a threshold 
SNR $>8$ in a single detector, and we use the fiducial O1 noise curve~\cite{fn1}. 
We calculate the face-on, overhead SNR for each binary directly from Eq.~2 of
\cite{Dominik2015}. We then calculate the luminosity
distance at which this binary would be detected with $\mbox{SNR}=8$. Note that
as the distance to the binary changes, the observer frame (redshifted) mass also
changes, and therefore calculating the horizon redshift requires an iterative
process. Once this has been calculated, we then determine the predicted
detection rates using Eq.~9 of~\cite{Dominik2015}. The effects of the antenna
power pattern are incorporated in the $p_{\rm det}$ term in this equation.

Estimate of fiducial aLIGO sensitivity during the 16-day GW150914 analysis
is shown in Figure~\ref{fig:RateComp}. We estimate the sensitivity to coalescing 
compact binaries using a reference O1 noise curve. We assume that both detectors 
operate with the fiducial O1 noise curve, which is the same sensitivity we adopted 
to calculate compact binary detection rates. For comparison, this model agrees 
reasonably well with the ``early-high'' sensitivity model~\cite{LVC2013}. Our 
expression is a 50th percentile upper limit, assuming no detections. The critical 
application of this expression is not related to its overall normalization; we 
are instead interested in its shape, which characterizes the strongly mass-dependent 
selection biases of LIGO searches.   

Using these inputs, our fiducial estimate of the advanced LIGO sensitivity during 
the first 16 days of O1 for a specific mass bin,
$\Delta M_{\rm i}$, is given by
\begin{eqnarray}
R_{{\rm D},\Delta M_{\rm i}, UL} = \frac{0.7}{V_{\Delta M_{\rm i}}T} ,
\end{eqnarray}
where $T$ is 16 days, corresponding to the analysis of GW150914~\cite{DiscoveryPaper}, 
and the volume 
$$
V_{\Delta M_{\rm i}} = (\Delta M_{\rm i})^{-1}\int dM \int \frac{dz}{1+z} \frac{dV}{dz} 
p_{\rm det}(<w,M)
$$
is the  sensitive volume averaged over the mass bin $\Delta M_{\rm i}$, and
$p_{\rm det}(<w,M)$ is the orientation-averaged detection 
probability~\cite{Dominik2015,Belczynski2015}.  
The function $p_{\rm det}(<w,M)$ depends on the coalescing binary redshifted mass 
through the maximum luminosity distance (``horizon distance'') at which a source could 
produce a response of SNR$>$8 in a single detector. To calculate this distance, we 
adopt the same model for inspiral, merger, and ringdown~\cite{Husa2015,Khan2015} used 
in the text to estimate compact binary detection rates. Extended Data 
Figure~\ref{fig:Horizon} shows our estimated horizon redshift as a function of the 
total redshifted binary merger mass for equal mass mergers.

\noindent
{\bf Code availability.}
We have opted not to release the population synthesis code {\tt StarTrack} 
used to generate binary populations for this study.

\clearpage

\renewcommand{\figurename}{Extended Data Figure}
\setcounter{figure}{0}

\renewcommand{\tablename}{Extended Data Table}
\setcounter{table}{0}

\vspace*{0cm}
\begin{table}
\begin{tabular}{cr|r|rrr}
Channel& Evolutionary sequence \hspace*{1.2cm} & all [$\%$] & high-$z$ & mid-$z$ & low-$z$\\
\hline \hline
BHBH1 &  MT1(2-1)\ \ \ BH1\ \ \ CE2(14-4;14-7)\ \ \ BH2 & 79.481 & 38.045 & 18.673 & 22.763 \\
BHBH2 &  MT1(4-1)\ \ \ BH1\ \ \ CE2(14-4;14-7)\ \ \ BH2 & 13.461 & 10.766 & 1.101 &  1.594 \\
BHBH3 &  MT1(4-4)\ \ \ CE2(4/7-4;7-7)\ \ \ BH1\ \ \ BH2 &  5.363 &  4.852 & 0.194 &  0.317 \\
Other &                        additional combinations  &  1.696 &  0.625 & 0.421 &  0.649 \\
\hline
\end{tabular}
\caption{\label{tab:EvolChan}{\bf Formation channels of massive BH-BH mergers (M1)}: 
The first two columns identify evolutionary sequences leading to the formation of 
BH-BH mergers with $M_{\rm tot,z}=54$--$73\msun$. The third column 
lists the formation efficiency. Last three columns list the formation efficiency of BH-BH 
progenitors born at $z>1.12$, $1.12<z<0.34$, $z<0.34$.
Notation: stable mass transfer (MT), common envelope (CE), BH formation (BH) initiated either 
by the primary star ($1$) or the secondary star ($2$). 
In parentheses we give the evolutionary stage of stars during MT/(pre;post)CE: main sequence 
($1$), Hertzsprung gap ($2$), core He-burning ($4$), helium star ($7$), BH ($14$), with the 
primary star listed first. 
}
\end{table}

\clearpage

\begin{figure}
\vspace*{-2.0cm}
\includegraphics[width=0.9\columnwidth]{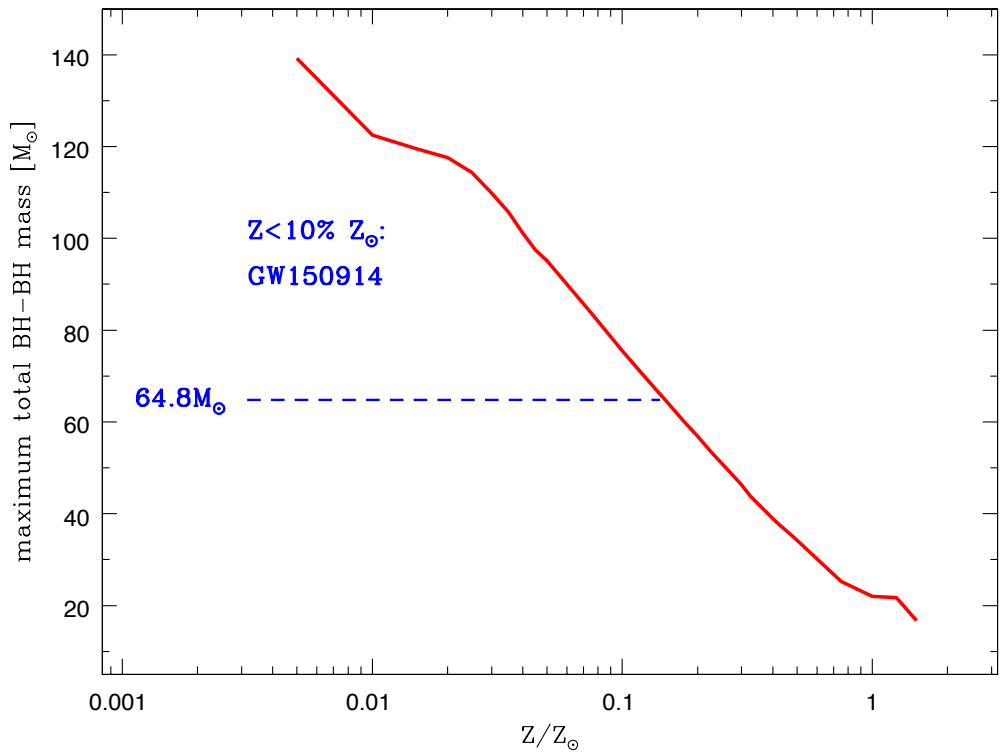}\vspace*{-4.5cm}\\
\caption{{\bf Maximum total mass of BH-BH mergers as a function of metallicity.} 
Binary stars at metallicities lower than $10\%$ solar can form BH-BH mergers more 
massive than $M_{\rm tot}=64.8\msun$. This suggests that GW150914 was formed in a 
low metallicity environment, assuming it is a product of classical isolated binary 
evolution. Note that the total {\em binary}\/ maximum BH-BH mass is not a simple 
sum of maximum BH masses resulting from {\em single} stellar evolution; this is a 
result of mass loss during the RLOF and CE evolution phases in the formation of 
massive BH-BH mergers (Fig.~\ref{fig:ExamplEvol}).} 
\label{fig:BHBHMass}
\end{figure}

\begin{figure}
\vspace*{-5.3cm}
\hspace*{-3.1cm}
\includegraphics[width=1.2\columnwidth]{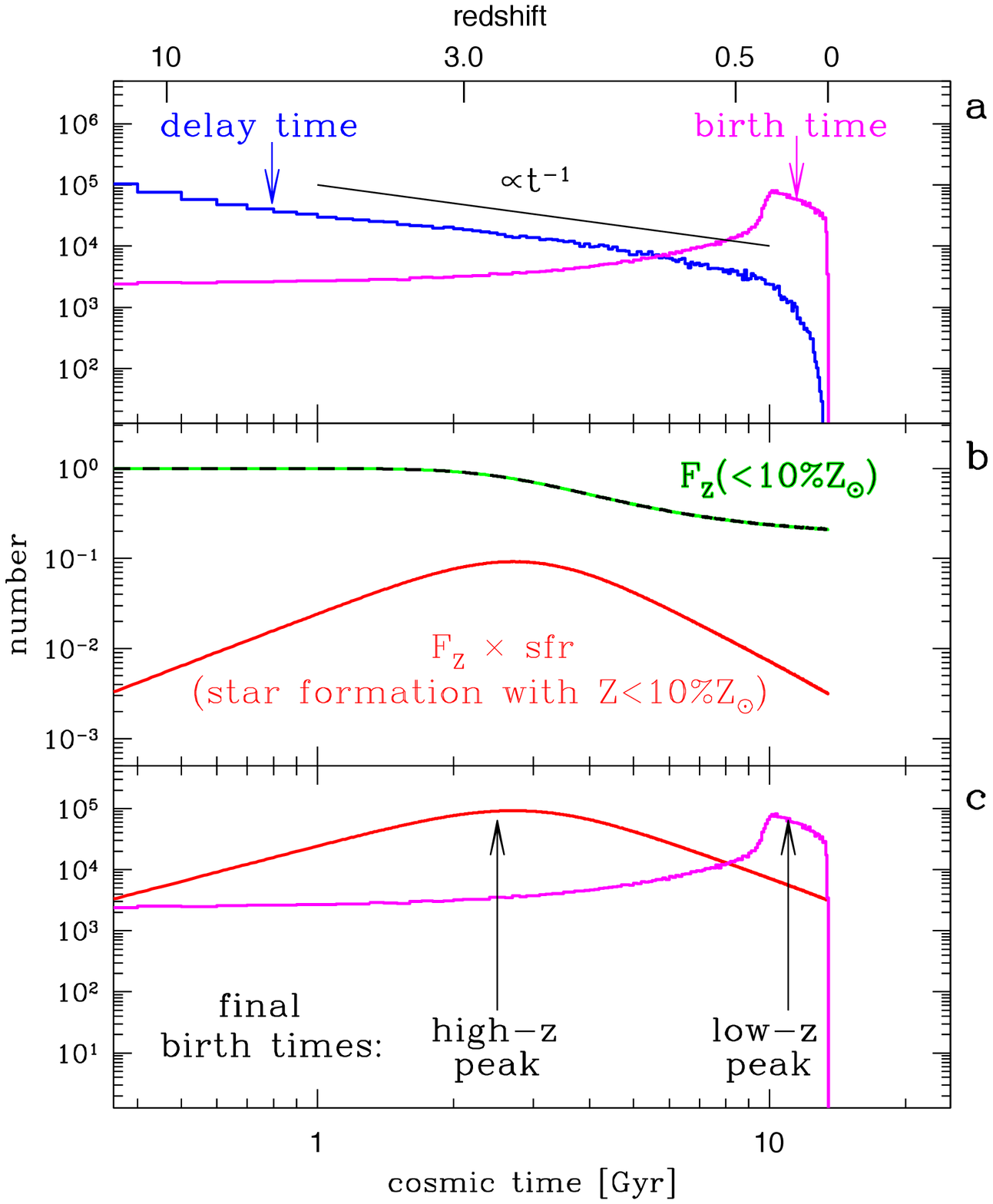}\vspace*{-5.5cm}\\
\caption{{\bf Emergence of bimodal birth time distribution.}
{\bf a,} 
Black hole binaries follow an intrinsic power-law delay time distribution 
($\propto t^{-1}$). 
The birth time ($t_{\rm birth}=t_{\rm merger}-t_{\rm delay}$) is inverted 
compared to the delay time distribution, with the spread caused by allowing the merger 
time ($t_{\rm merger}$) to fall anywhere within the LIGO O1 horizon: $z=0$--$0.7$; 
this generates a peak corresponding to BH-BH progenitors born late with short delay 
times.
{\bf b,} 
Massive BH-BH binaries are formed only by low-metallicity stars ($Z<10\%\zsun$). 
The fraction of all stars that form at such low-Z ($F_{\rm Z}$) decreases with cosmic 
time making low-Z star formation [$\msun$ Mpc$^{-3}$ yr$^{-1}$] peak at early cosmic 
time.      
{\bf c,} 
Final birth time distribution for massive BH-BH mergers is a convolution of 
the intrinsic birth times with the low metallicity star formation rate.}
\label{fig:Emergence}
\end{figure}

\begin{figure}
\vspace*{-4cm}
\includegraphics[width=1.0\columnwidth]{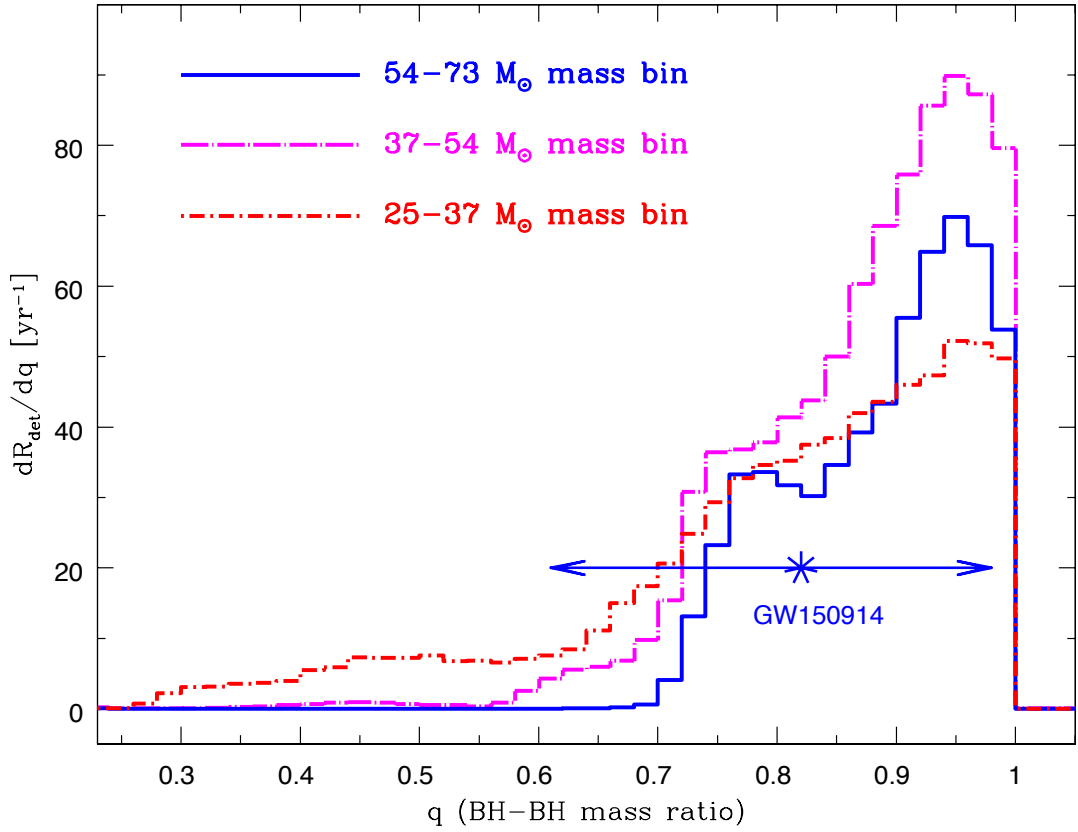}\vspace*{-4.5cm}\\
\caption{{\bf Predicted distribution of BH-BH merger mass ratios}.  
Standard model (M1) detector frame mass ratio is shown. BH-BH binaries prefer 
mass ratios of $q\gtrsim 0.7$, with a prominent peak near comparable-mass 
systems. GW150914 with $q=0.82_{-0.21}^{+0.16}$ ($90\%$ credible range) and with 
total redshifted mass of $M_{\rm tot,z}=70.5\msun$ falls within the expected 
region.}
\label{fig:MassRatio}
\end{figure}

\begin{figure}
\vspace*{-4.0cm}
\includegraphics[width=1.0\columnwidth]{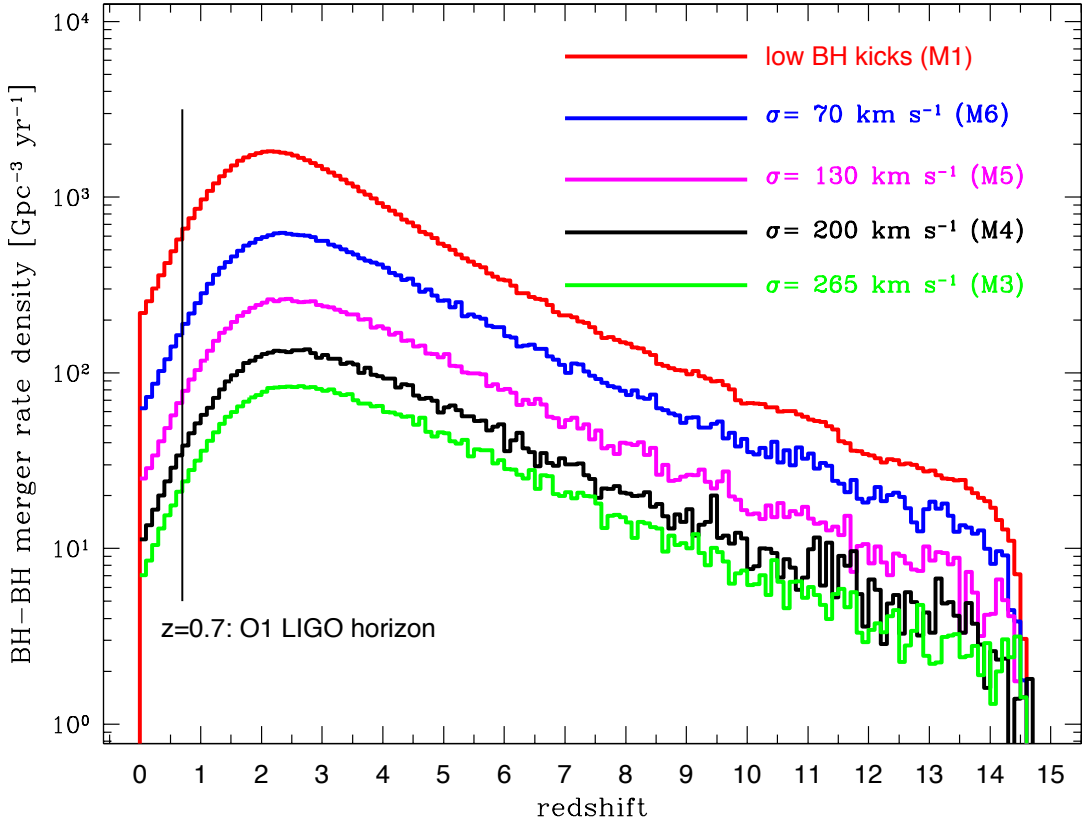}\vspace*{-4.5cm}\\
\caption{
{\bf Source frame merger rate density for BH-BH binaries as a function of redshift.}
The red line shows the results from our standard model (M1); 
in this model massive BHs do not get natal kicks. A sequence of models with increasing 
BH natal kicks (models M6, M5, M4, M3) is shown. The rate density decreases with 
increasing natal kick. The local merger rate density ($z<0.1$) changes from $218\gpy$ 
(M1), to $63\gpy$ (M6), $25\gpy$ (M5), $11\gpy$ (M4), $6.6\gpy$ (M3). The LIGO 
estimate ($2$--$400\gpy$) encompasses all of these models. We mark the O1 LIGO 
detection horizon ($z=0.7$; see Extended Data Fig.~\ref{fig:Horizon}).} 
\label{fig:IntrRate}
\end{figure}

\begin{figure}
\vspace*{-4.0cm}
\includegraphics[width=1.0\columnwidth]{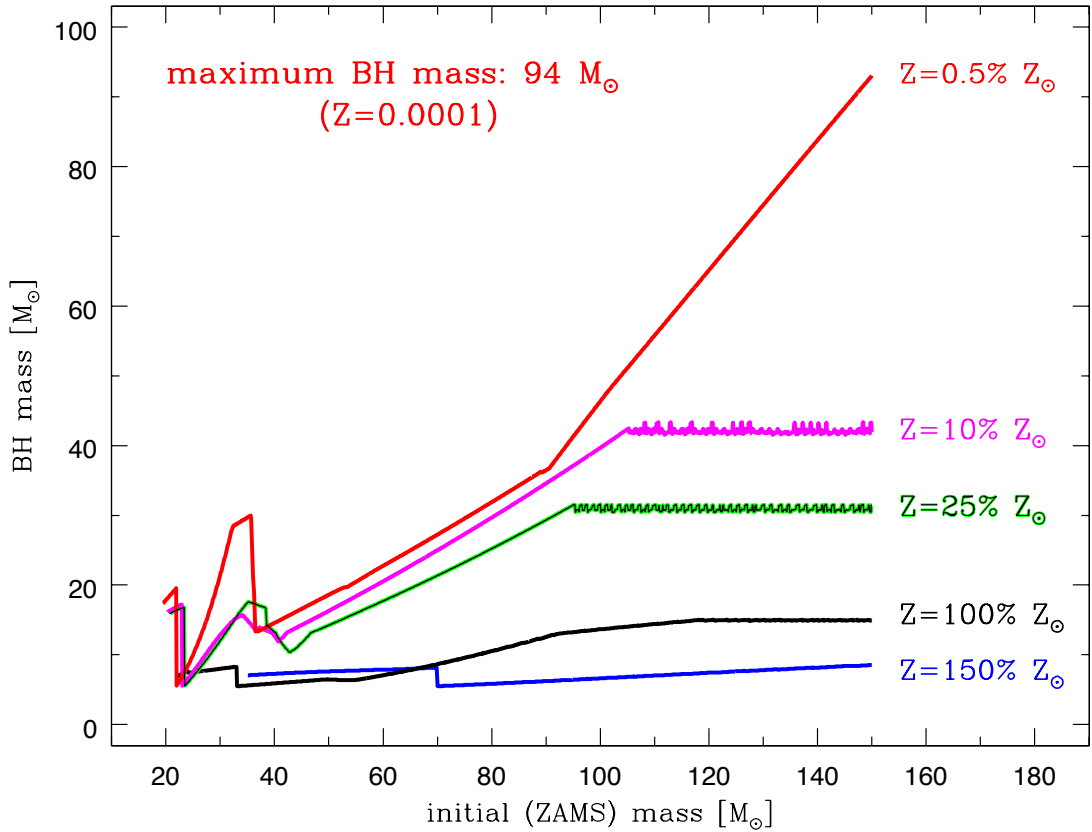}\vspace*{-4.5cm}\\
\caption{{\bf BH mass as a function of initial star mass, for a range of metallicities.} 
These results show calculations for single star evolution with no binary interactions. 
Our updated models of BH formation show a general increase of BH mass with initial 
progenitor star mass. There is strong dependence of BH mass on the chemical 
composition of the progenitor. For example, the maximum BH mass increases from 
$10$--$15\msun$ for high metallicity progenitors ($Z=150$--$100\%\zsun$) to 
$94\msun$ for low metallicity progenitors ($Z=0.5\%\zsun$). Note that the formation 
of a single $30\msun$ BH requires metallicity of $Z \leq 25\%\zsun$.}
\label{fig:BHMassRel}
\end{figure}

\begin{figure}
\vspace*{-4.0cm}
\includegraphics[width=1.0\columnwidth]{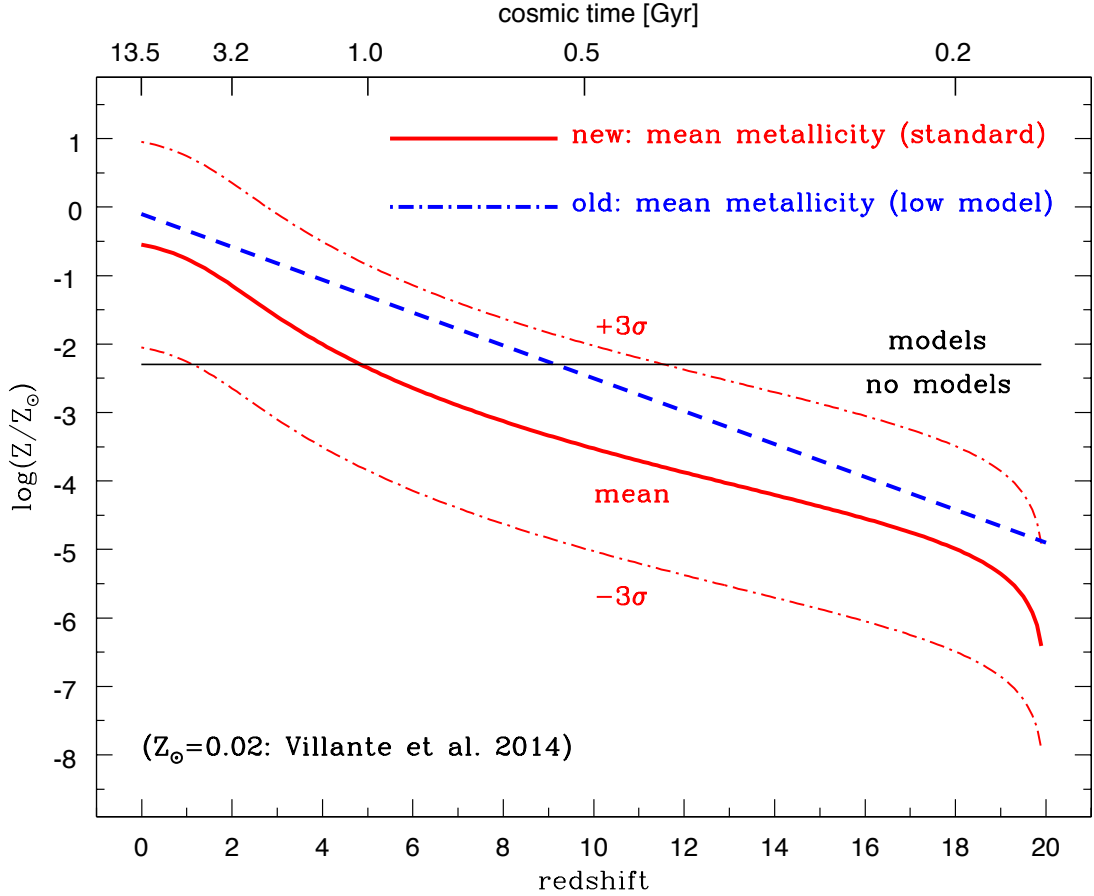}\vspace*{-4.5cm}\\
\caption{{\bf Mean metallicity evolution of the Universe with redshift.} 
It is assumed that at each redshift the metallicity distribution is log-normal 
with a standard deviation $\sigma=0.5$dex. The blue line denotes the mean 
metallicity evolution adopted in previous studies. The new relation generates 
more low metallicity stars at all redshifts. We mark the line above which we can 
make predictions ($\log(Z/Z_\odot)=-2.3$; solar metallicity
$Z_\odot=0.02$~\cite{Villante2014}) based on actual evolutionary stellar 
models adopted in our calculations. Below this line we assume that stars produce 
BH-BH mergers in the same way as in the case of our lowest available model. }
\label{fig:MetEvol}
\end{figure}

\begin{figure}
\vspace*{-4.0cm}
\hspace*{-0.3cm}
\includegraphics[width=1.0\columnwidth]{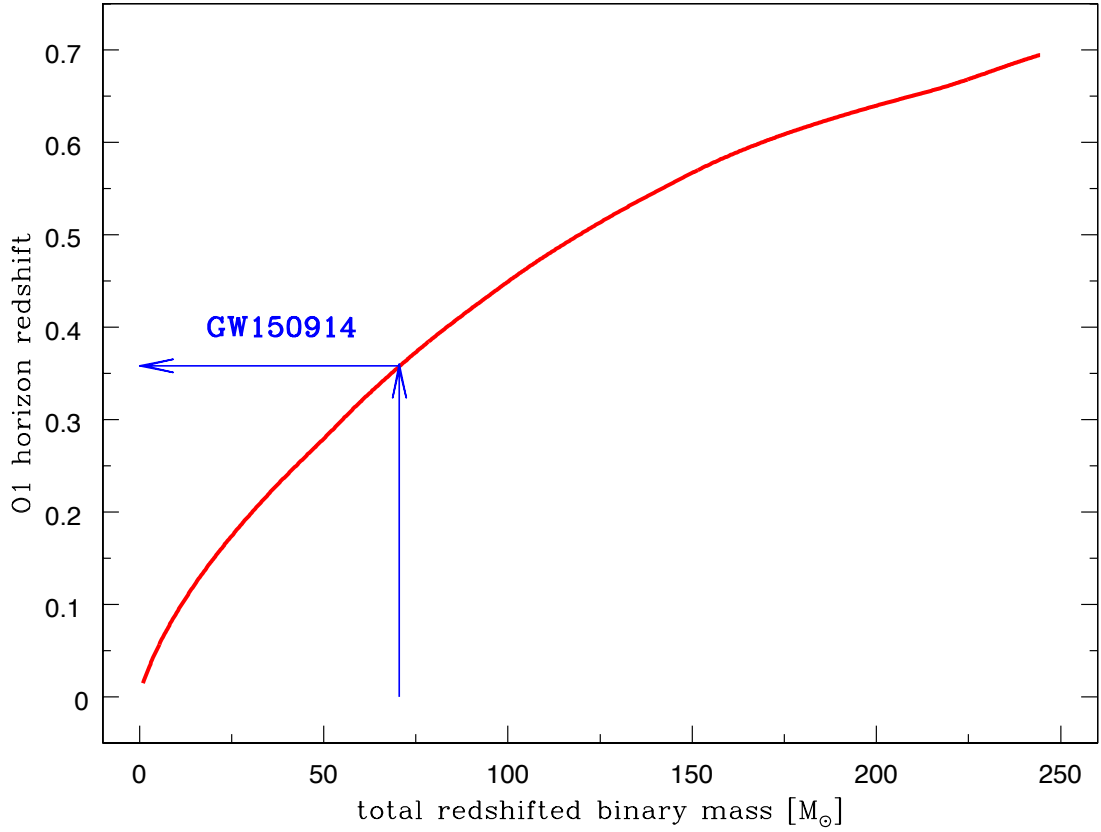}\vspace*{-4.5cm}\\
\caption{{\bf Horizon redshift for the first advanced LIGO observational run (O1).}
Horizon is given as a function of the total redshifted binary merger mass 
(assuming equal-mass mergers). For the highest mass mergers found in our 
simulations ($M_{\rm tot,z} = 240\msun$) the horizon redshift is $z_{\rm hor}=0.7$. 
For GW150914 ($M_{\rm tot,z}=70.5\msun$) the horizon redshift is $z_{\rm hor}=0.36$.}
\label{fig:Horizon}
\end{figure}

\end{document}